\documentclass[aps,prd,twocolumn,amsmath,amssymb,amsfonts,nofootinbib,superscriptaddress,altaffilletter]{revtex4-2}
\usepackage{graphicx}
\usepackage{soul}
\usepackage{cellspace}
\usepackage{amssymb}
\usepackage{amsmath}
\usepackage{longtable}
\usepackage{tabu}
\usepackage{color}
\usepackage{etoolbox}
\usepackage{supertabular}
\usepackage[caption=false]{subfig}
\usepackage{savesym}
\savesymbol{tablenum}
\usepackage{siunitx}
\restoresymbol{SIX}{tablenum}
\usepackage{dcolumn}
\usepackage{ wasysym }
\usepackage{ulem}
\usepackage{amsmath}

\providecommand{\gt}{>}

\newcommand{\NRDeClPercent}{\num{99.99}}

\def\threshIter{\Gamma_\mathrm{S}}
\def\threshLow{\Gamma_\mathrm{L}}

\newcommand{\mathO}[1]{\mathcal{O}\left(#1\right)}

\newcommand{\occ}{N_{occ}{}}
\newcommand{\occThresh}{\occ_{,\mathrm{min}}}

\newcommand{\nbins}{N_{bins}{}}
\newcommand{\nbinsThresh}{N_{\mathrm{bins,min}}}

\newcommand{\NR}{\mathrm{NR}}

\newcommand{\eclx}{x_\mathrm{ecl}}
\newcommand{\ecly}{y_\mathrm{ecl}}

\newcommand{\binspacing}{\delta b}
\newcommand{\binspacingf}{\delta f}
\newcommand{\binspacingfdot}{\delta \dot{f}}
\newcommand{\binspacingfddot}{\delta \ddot{f}}
\newcommand{\binspacingeclx}{\delta x_{\mathrm{ecl}}}
\newcommand{\binspacingecly}{\delta y_{\mathrm{ecl}}}

\newcommand{\Freq}{f}
\newcommand{\fdot}{{\dot{\Freq}}}

\newcommand{\Gauss}{\mathrm{\MakeUppercase{G}}}
\newcommand{\Signal}{{\mathrm{\MakeUppercase{S}}}}
\newcommand{\Line}{{\mathrm{\MakeUppercase{L}}}}
\newcommand{\Transient}{{\mathrm{t\MakeUppercase{L}}}}

\newcommand{\NoisetL}{{\Gauss\Line\Transient}}

\newcommand{\BSNtsc}{{\hat\beta}_{{\Signal/\NoisetL}}}	

\newcommand{\SDNinety}{\mathcal{D}^{90\%}}

\newcommand{\taxijump}{N^j}
\newcommand{\taxiconstr}{N_c}

\begin{document}
	\title[Density Clustering Algorithm]{Density-clustering of continuous gravitational wave candidates from large surveys}
	
	\author{B. Steltner}
	\email{benjamin.steltner@aei.mpg.de}
	\affiliation{Max Planck Institute for Gravitational Physics (Albert Einstein Institute), Callinstrasse 38, 30167 Hannover, Germany}
	\affiliation{Leibniz Universit\"at Hannover, D-30167 Hannover, Germany}
	
	\author{T. Menne}
	\affiliation{Max Planck Institute for Gravitational Physics (Albert Einstein Institute), Callinstrasse 38, 30167 Hannover, Germany}
	\affiliation{Leibniz Universit\"at Hannover, D-30167 Hannover, Germany}
	
	\author{M. A. Papa}
	\affiliation{Max Planck Institute for Gravitational Physics (Albert Einstein Institute), Callinstrasse 38, 30167 Hannover, Germany}
	\affiliation{Leibniz Universit\"at Hannover, D-30167 Hannover, Germany}
	\affiliation{University of Wisconsin Milwaukee, 3135 N Maryland Ave, Milwaukee, WI 53211, USA}

	\author{H.-B. Eggenstein}
	\affiliation{Max Planck Institute for Gravitational Physics (Albert Einstein Institute), Callinstrasse 38, 30167 Hannover, Germany}
	\affiliation{Leibniz Universit\"at Hannover, D-30167 Hannover, Germany}
	
	
	\begin{abstract}
		Searches for continuous gravitational waves target nearly monochromatic gravitational wave emission from e.g. non-axysmmetric fast-spinning neutron stars. Broad surveys often require to explicitly search for a very large number of different waveforms, easily exceeding $\sim10^{17}$ templates.  In such cases, for practical reasons, only the top, say $\sim10^{10}$, results are saved and followed-up through a hierarchy of stages. Most of these candidates are not completely independent of neighbouring ones, but arise due to some common cause: a fluctuation, a signal or a disturbance. By judiciously clustering together candidates stemming from the same root cause, the subsequent follow-ups become more effective. A number of clustering algorithms have been employed in past searches based on iteratively finding symmetric and compact over-densities around candidates with high detection statistic values. The new clustering method presented in this paper  is a significant improvement over previous methods: it is agnostic about the shape of the over-densities, is very efficient and it is effective: at a very high detection efficiency, it has a noise rejection of $\NRDeClPercent\%$ , is capable of clustering two orders of magnitude more candidates than attainable before and, at fixed sensitivity it enables more than a factor of 30 faster follow-ups. We also demonstrate how to optimally choose the clustering parameters. 
	\end{abstract}
	
	\maketitle
	
	\section{Introduction}
	\label{sec:Intro}
	Continuous gravitational waves are long-lasting signals that may come from fast-spinning non-axisymmetric neutron stars, unstable r-modes \cite{Lasky:2015uia,Owen:1998xg}, the fast inspiral of dark-matter objects \cite{Horowitz:2019aim,Horowitz:2019pru} or emission from clouds of axion-like particles around black holes \cite{Arvanitaki:2014wva,Zhu:2020tht}.
	
	Independently of the emission mechanism, the expected signal is a nearly monochromatic wave that due to the relative motion between the source and the detector, is frequency- and amplitude- modulated. The signal shape is described by a frequency and its derivatives, the source position in the sky, the signal amplitude, the polarization angle and the source inclination with respect to the line of sight. 
	
	Searches for continuous gravitational wave signals typically use template banks for frequency, frequency derivatives and source position only, because the remaining parameters can be analytically maximized over, and do not need to be searched for explicitly. The parameter space of broad surveys grows quickly with the observation span, and for observations lasting months $\sim{10^{17}}$ template waveforms need to be considered.  
	
	For template banks that are this big, typically only the top results are saved -- say the $\sim{10^{10}}$ results with the highest detection statistic values. These are then considered for further follow-up investigations \cite{Papa:2016cwb, Abbott:2017pqa, Steltner:2020hfd}. Even though at this stage most of the results are not statistically significant, because they will be subject to more inspections, they are referred to as ``candidates".
	
	To reduce the loss in signal-to-noise ratio due to template-signal waveform mismatch, our templates have a high overlap and thus are not independent. Therefore a loud signal or disturbance triggers multiple templates, generating a high number of candidates close to each other in parameter space. Following-up each one independently would result in a waste of resources.
	
	Clustering is an important step in the post-processing of the results that organizes and reduces the $\sim{10^{10}}$ candidates to a more useful set of $\approx$ independent $\sim{10^{6}}$ candidates. 
	
	While the clustering algorithm details vary, the core is to find candidates likely due to the same root cause, bundle (\emph{cluster}) them and consider them as a single entity in follow-up studies. 
	
	Each cluster is represented by the parameters of the so-called \emph{seed} candidate and by a \emph{containment region}. The latter  measures how far from the seed associated with a signal, the true signal parameters are. In follow-up studies the entire containment region around each seed is surveyed. The containment region is the same for all seeds and it is determined statistically, such that it holds for a very large fraction ($\gt 99\%$) of signals, across the parameter space.
	
	It has also been observed that a threshold on the minimum number of candidates in a cluster is effective at discarding noise-clusters. With a fixed computing budget for follow-ups, fewer candidates means that freed-up computational capacity can be used on additional, lower significance candidates which translates in deeper and more sensitive searches.
	
	Clustering is a crucial step in the analysis of the results of broad continuous wave surveys with very large template banks, such as those carried out on the volunteer computing project Einstein@Home\footnote{\href{https://www.einsteinathome.org/}{www.einsteinathome.org/}} \cite{Boinc1, Boinc2, Boinc3}. For this reason clustering procedures have been in use for a long time: One of the first non-trivial clustering procedures is box-clustering \cite{Aasi:2013jya,Behnke:2014tma}, which dates back to nearly a decade ago. More recently a more flexible adaptive clustering technique has been used \cite{Singh:2017kss} which however does not converge fast enough when used on many data points. This is a significant drawback, as we want to set lower thresholds, which means considering more candidates in the follow-ups. Attempts to use machine-learning for clustering have been successful for directed searches, but not for all-sky searches \cite{Beheshtipour:2020zhb, Beheshtipour:2020nko}.

	We present here the new \emph{Density Clustering} algorithm, able to process orders of magnitude more candidates than previous clustering strategies at comparable, if not lower, computing cost. We show how to choose the clustering parameters, and demonstrate its performance on real data. We concentrate on clustering results from very large template banks -- with over $10^{16}$ points -- and hence refer to the Einstein@Home results. This method can also be employed in less challenging environments.
	
	The paper is organized as follows: In Section \ref{sec:data} we describe the input data; in Section \ref{sec:deCl} the method itself; in Section \ref{sec:hyperparameter} the choice of the clustering parameters; in Section \ref{sec:implementation} the implementation; in Section \ref{sec:performance} the method is compared with Adaptive Clustering under realistic conditions, i.e. by applying it to the data of the Stage 0 results of the Einstein@Home all-sky search for continuous gravitational waves in Advanced LIGO data of the second observation run (O2) \cite{Steltner:2020hfd, Vallisneri:2014vxa}.
	
	\section{Input data to clustering}
	\label{sec:data}
	
	Clustering works on a set of candidates, i.e. selected results from a search. A candidate is described by the values of the template that produced the detection statistic result, and the detection statistic result. For an all-sky search including up to second-order spin-down parameters, a generic candidate $i$ is of the form 
	\begin{equation}
		\left(f_i, \dot{f}_i, \ddot{f}_i, \alpha_i, \delta_i, \chi_i=\BSNtsc{}_i\right), 
	\end{equation}
	where $f$ indicates the signal-template frequency, $\alpha,\delta$ the source sky position and $\chi$ the detection statistic. Consistently with the Einstein@Home searches we have indicated $\BSNtsc{}$ the line-and-transient-line robust statistic  \cite{Keitel:2015ova} as the detection statistic of choice. We will illustrate clustering for these 5 dimensions; fewer or more dimensions are treated analogously.
	
	\begin{figure*}[htbp]
		\centering
		\includegraphics[width=\textwidth]{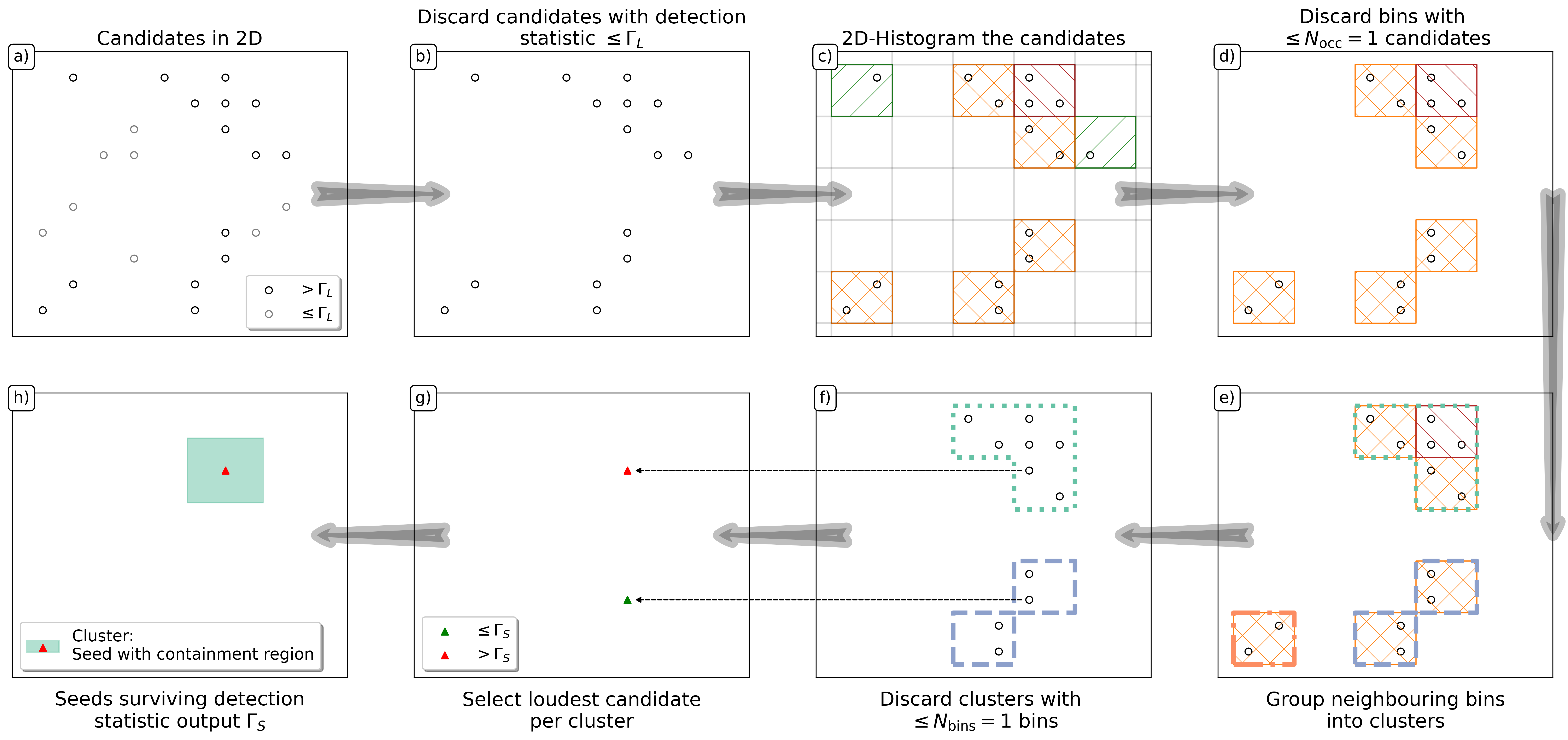}
		\caption{Schematic illustration of the main steps of Density Clustering}
	\label{fig:MockExample}
\end{figure*}

Since continuous waves are modulated by the Earth's rotation and orbit around the Sun, the sky grids are set-up in sky coordinates projected on the ecliptic plane,$\eclx, \ecly$. Therefore for clustering we convert for the candidates $(\alpha_i, \delta_i) \rightarrow (\eclx{}_i, \ecly{}_i)$ -- see Eq.s~(14-15) in \cite{Singh:2017kss} for the conversion between $(\alpha, \delta) \rightarrow (\eclx, \ecly)$. 

The sky grids are approximately uniform hexagonal grids on the ecliptic plane and are defined by the hexagon edge length $d$:
\begin{equation}
	d(m_{\text{sky}})={1\over f}{	{\sqrt{ m_{\text{sky}} } }	\over {\pi \tau_{E}}} ,\label{eq:skyGridSpacing}
\end{equation}
with $\tau_{E}\simeq0.021$ s being half of the light travel-time across the Earth and $m_{\text{sky}}$ a constant which controls the resolution of the sky grid \cite{Steltner:2020hfd}. From Eq.~\ref{eq:skyGridSpacing} it is clear that the sky-grid density increases with frequency $f$.

\section{Density Clustering}
\label{sec:deCl}
We bin the parameter space in equally-spaced cells of size 
\begin{equation}
	\binspacing = (\binspacingf, \binspacingfdot, \binspacingfddot, \binspacingeclx, \binspacingecly)
\end{equation}
in each dimension. The $\binspacingf, \binspacingfdot, \binspacingfddot$ are each an integer multiple of the search grid spacing.  The sky grid has a hexagonal tilling, so the square tiling of the bins above does not match it. The bins are usually chosen to be large enough that this does not matter and the square covering greatly simplifies the binning and the identification of neighbouring bins. The bin size is always a multiple of the hexagon side, so the bins shrink with increasing frequency as the sky-grid pixels, keeping the average number of candidates per bin the same.

We only consider candidates with detection statistic values above a threshold $\Gamma_L$. 
In each bin $j$ we count the number of candidates $\occ{}_{,j}$ with parameters in that bin. Bins with $\occ{}_{,j}\leq\occThresh$ are discarded. $\occThresh$ is one of the clustering parameters and its optimal value depends on the search set-up and on the bin size.

Among the surviving bins, we cluster together nearby ones, to create a cluster. The basic notion of vicinity is controlled by two parameters: $\taxijump$ and $\taxiconstr$. A bin $b_a$ is a neighbour of bin $b_c$ if the distances $k^j$ in integer bin spacings 
\begin{equation}
	b_a - b_c = \left(k^{1} \binspacingf, k^{2} \binspacingfdot, k^{3} \binspacingfddot, k^{4} \binspacingeclx, k^{5} \binspacingecly\right)
\end{equation}
satisfy the following conditions:
\begin{equation}
	\begin{cases} k^{j} \leq \taxijump ~~~{\textrm{with}} ~j=1,..,M \\
		\sum_{j=1}^M{k^{j}} \leq \taxiconstr ,
	\end{cases}
\end{equation}
where $M$ is the number of dimensions.  The first condition sets the maximum distance in every dimension, whereas the second condition sets an overall maximum distance.  With $M=3$, $\taxiconstr = 1$ means that the two nearby bins have to share a face, $\taxiconstr = 2$ that they have to share an edge and $\taxiconstr = 3$ that they have to share a vertex. Default values are $\taxijump=1$, equal for all $j$, and $\taxiconstr=M$. 

Among the clusters from the previous step, we remove the ones with too few bins: $\nbins \leq \nbinsThresh$.

For each remaining cluster a representative candidate becomes the seed. The seed is by default the candidate with the highest detection statistic value (the loudest) of all candidates in the cluster. In noisier data it may make sense to look at the loudest candidate in the bin with the most candidates (densest bin) or the loudest candidate in the bin with the highest sum over all detection statistic values of the candidates within that bin (loudest bin). 

Finally all clusters with a seed with detection statistic value smaller than $\threshIter$ are discarded.

An additional parameter can be used to mitigate binning effects: an overdensity of candidates may not be perfectly contained within one bin, but may extend across bin boundaries. 
For faint signals with just enough candidates to surpass the occupancy threshold $\occThresh$, this effect can make the difference between recovering a signal or not. Boundary effects can be partly mitigated by smoothing over bins, e.g. adding bin counts over neighbouring bins or adding bin counts weighted with a Gaussian kernel. The overall impact of using smoothing procedures should be evaluated within the general framework of choosing the optimal clustering parameters, as described in the next Section, but we will not explicitly consider it here.

\begin{table*}[th]
	\begin{tabular}{|l|l|}
		\hline
		\hline
		Parameter & Function \\	\hline
		\hline
		Input-Threshold $\threshLow$ & Discards candidates with detection statistic $\leq \threshLow$. Filters input-candidates \\	\hline
		Bin sizes  $\binspacing$ & Binning \\	\hline
		Smoothing & Smooth histogram or not \\	\hline
		Occupancy-threshold $\occThresh$ & Discard bins with $\occ \leq \occThresh $ candidates \\	\hline
		Neighbour-criterion, $\taxijump$ and $\taxiconstr$ & Defines what a neighbour is \\	\hline
		Cluster-size-threshold $\nbins$ & Discard clusters with $\nbins \leq \nbinsThresh$ bins \\	\hline
		Seed criterion & Loudest candidate in cluster, loudest in most-populated bin \\	
		&  or in bin with highest average detection statistic\\ \hline
		Output-Threshold $\threshIter$ & Discards cluster whose seed has detection statistic $\leq \threshIter$. Reduces false alarms \\	\hline
		\hline
	\end{tabular}
	\caption{Parameters of Density Clustering in the order that they are employed}
	\label{tab:parameters}
\end{table*}

\section{Choosing the parameters of the clustering procedure}\label{sec:hyperparameter}

A number of parameters define the Density Clustering algorithm, and they are summarized in Table \ref{tab:parameters}. We choose the parameters such that at fixed computational cost for the follow-up of the resulting seeds, the sensitivity of the clustering procedure is maximized.

The sensitivity of the clustering procedure is measured by the gravitational wave signal amplitude $h_0^{90\%}$ at which the detection efficiency $\epsilon$ of the clustering procedure is 90\%, for signals with parameters in the search range. $h_0^{90\%}$ scales with the amplitude spectral density of the noise $\sqrt{S_h(f)}$, so we will consider the quantity $\SDNinety=\sqrt{S_h(f)}/ {h_0}$, instead, that does not depend on frequency.  $\mathcal{D}$ is also known as the sensitivity depth \cite{Behnke:2014tma}. 

Since there is no way to predict the detection efficiency of the clustering procedure, we measure it with a Monte Carlo, where we add fake signals to the noise with amplitudes corresponding to a given value of $\mathcal{D}$. For each signal we perform a search like the one whose results we wish to cluster, cluster the results and produce seeds. If one of the seeds comes from the added signal, we consider the signal detected. The fraction of detected signals to total signals gives the detection efficiency at that sensitivity depth: $\epsilon(\mathcal{D})$. $\SDNinety$ is then
\begin{equation}
	\epsilon(\SDNinety)=90\%,
	\label{eq:SD90}
\end{equation}
and $\SDNinety$ measures the sensitivity of the clustering procedure: the higher is $\SDNinety$, the higher is the sensitivity.

\begin{figure}[htb]
	\includegraphics[width=\columnwidth]{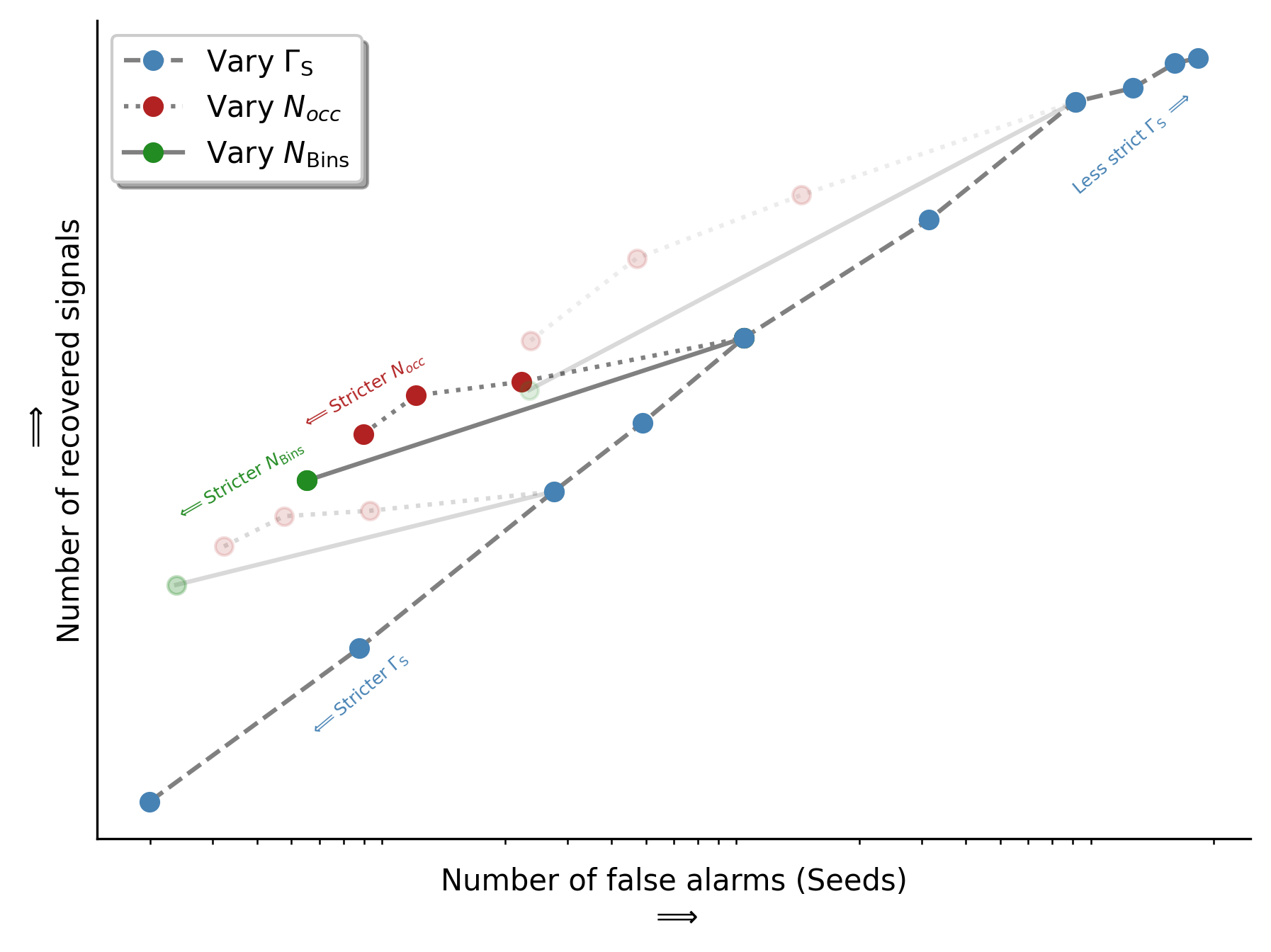}
	\caption{We show how the number of false alarms (in log-scale) and the number of detected signals changes while varying just a single  clustering parameter. Along the dashed line the output threshold $\threshIter \in [-4 ,10]$ varies. The stricter (higher) $\threshIter$ is, the fewer signals are recovered and the fewer are the false alarms. Similarly stricter $\occThresh$ (dotted line) or $\nbinsThresh$ (solid line) reduce the number of false alarms. The number of recovered signals is also reduced, but less than by using a higher $\threshIter$. It is often beneficial to allow lower significance candidates (lower $\threshIter$) and veto more aggressively based on $\occThresh$ and $\nbinsThresh$, instead.}
	\label{fig:paramChanges}
\end{figure}

We consider different clustering set-ups corresponding to different choices of clustering parameters. For each we estimate 
\begin{itemize}
	\item $\SDNinety$
	\item the containment region (see Section \ref{sec:Intro}).
	\item the false alarm rate. This is done by running the clustering on a sub-set on the search results, at different frequencies. Since we operate in the regime of very rare signals, we take this as a measure of the false alarm.
	\label{list:opt}
\end{itemize}
Figure~\ref{fig:paramChanges} shows how the detection efficiency and the false alarm rate change as different clustering parameters vary. 

The number of seeds and the containment region are used to estimate the computing cost of the follow-up. In principle one could optimize the follow-up search setup for each clustering setup. This would however be extremely expensive and experience has shown that a set-up choice guided by the sensitivity gain with respect to the previous stage, at accessible computing cost, lands a choice not significantly far from optimum. So we assume here that the follow-up set-up is fixed.  

\begin{figure}[htbp]
	\includegraphics[width=\columnwidth]{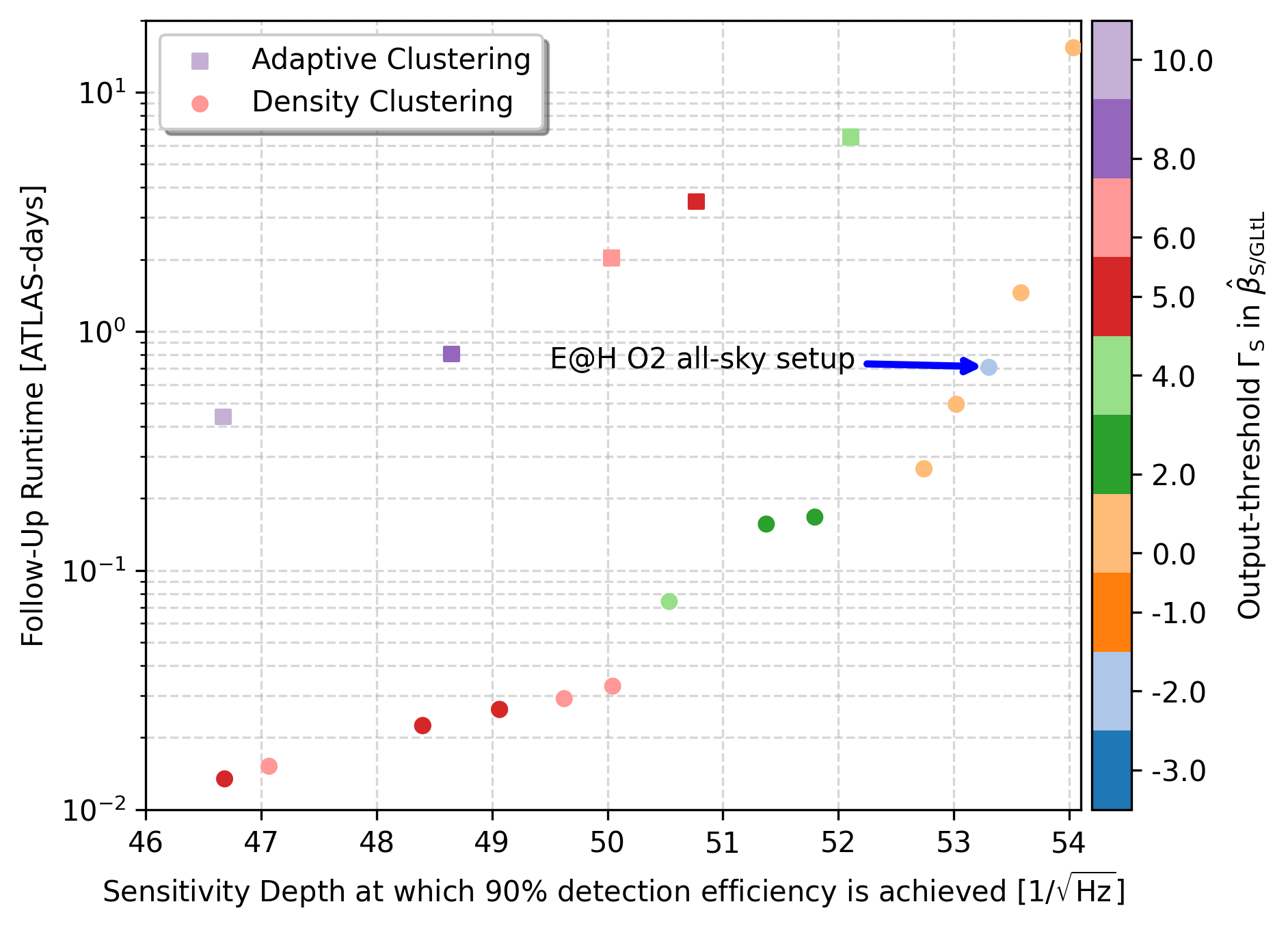}
	\caption{Performance comparison between the previous clustering method, Adaptive Clustering, and Density Clustering. Each point represents a different clustering set-up, used on the results of the Einstein@Home search \cite{Steltner:2020hfd}. To avoid excessive clutter we do not show all considered set-ups, but rather only those with runtime close to the smallest runtime at each $\SDNinety$. The color encodes the $\Gamma_S$ threshold parameter value. The arrow indicates the density clustering set-up chosen for the follow-up analysis reported in \cite{Steltner:2020hfd}.
	}
	\label{fig:SD90vsRuntime}
\end{figure}

We can now identify the clustering set-up that yields the highest $\SDNinety$, within the computing budget. This is illustrated in Figure \ref{fig:SD90vsRuntime} for the results of the Stage-0 Einstein@Home search \cite{Steltner:2020hfd}. 

\section{Implementation}\label{sec:implementation}

In the previous Section we have described how the optimal combination of clustering parameters is identified. As we have seen, this requires a Monte Carlo in order to measure the false alarm and 90\% detection-efficiency signal-amplitude $\SDNinety$, for every clustering set-up. 

For each setup we cluster $\gtrsim 2000$ result-files corresponding to data with different fake signals -- this is to determine $\SDNinety$. We cluster $\gtrsim 500$ search result-files with no fake signals, in order to estimate the false alarm. These operations can be quite time-consuming, so we describe here how to reduce the computing cost of this step. 

Einstein@Home search results typically come in files that cover a 50 mHz range of template frequencies, with size varying between a few MB to few GB, due to the different sky resolutions in the range $50-600$ Hz. Each clustering instance uses as input one of these 50 mHz results-files. Since the time to cluster is $\lll$ the time it takes to load such a file, it is faster to load a results-file, keep it in memory, and test different clustering set-ups.

Further savings are obtained by re-using intermediate results: 
\begin{itemize}
	\item we compute a histogram for a choice of $\threshLow$ and $\binspacing$, and re-use it to produce the bin-counts for different values of $\occThresh$  
	\item similarly, for a choice of $\threshLow$, $\binspacing$ and $\occThresh$ from the bin-counts we produce different clusters for different values of $\nbinsThresh$
	\item for each cluster different seeds are produced, based on different seed-selection criteria, e.g. the loudest cluster candidate, the loudest in the densest bin, the loudest in the bin with the highest average detection statistic (we call this the ``loudest bin"). By default this step is not performed and the loudest cluster candidate is directly considered as the ultimate cluster seed.
	\item finally, each seed, and with it the whole cluster, may be discarded depending on the value of $\threshIter$
\end{itemize}
With this scheme, testing a single clustering set-up costs (on average, over many set-ups) just under a second, with more than half  the time spent on the initial histogram and clustering. In order to reduce memory usage, the candidates are internally addressed only by an id. For thresholding on $\threshIter$ and for computing the containment region, the actual seed parameters must be retrieved. This operation accounts for another $20\%$ of the computing time. The remaining time is due to fluctuations in these estimates due to varying number of seeds and the initial overhead.

Given the computing-load profile described above, we parallelise the work among different independent processors, with each processor working only with a single results file and several ($\threshLow, \binspacing$)-combinations. Say we have $2500$ result-files, $1000$ ($\threshLow, \binspacing$) combinations and $1000$ combinations of the remaining parameters, each processors analyses 100 ($\threshLow, \binspacing$)-combinations, exhausting all $1000$ combinations of the remaining parameters. Hence, with $10$ processes per result file, $25000$ processes are spawned in total.

Using the large-capacity and fast-loading hdf5 and FITS file formats, and a HDD-raid configuration results-file server, testing a single ($\threshLow, \binspacing$)-combination and all $1000$ combinations of the remaining parameters, takes $\approx\SI{0.26}{\hour}$. Thus one processor exhausting 100 ($\threshLow, \binspacing$)-combinations takes $\approx$ a day. On the ATLAS cluster\footnote{ATLAS is the super-computer cluster at the MPI for Gravitational Physics in Hannover: \texttt{https://www.atlas.aei.uni-hannover.de/}} using $25000$ parallel processes the full testing of $1000\times 1000$ set-ups is carried out in a day.

\section{Performance on E@H O2 all-sky}\label{sec:performance}

We compare our Density Clustering with the Adaptive Clustering \cite{Singh:2017kss} on the results of the Stage-0 Einstein@Home O2 all-sky search \cite{Steltner:2020hfd}.

We characterize the detection efficiency on a set of $\sim 2900$ fake signals from the target source population of the search: signals with spin-frequencies uniformly distributed; spin-downs log-uniform distributed and all other parameters distributed uniformly: orientation $\cos\iota\in [-1,1)$, polarization angle $\psi\le\left|\pi/4\right|$, sky position $0\leq\alpha\le 2\pi$ and $-1 \leq\sin{\delta}\leq1$. The signal amplitude $h_0$ ranges from loud to faint signals with $\sim1000$ signals too faint to be detectable by either method. 

The results of the procedure described in the previous Section in order to identify the optimal Density Clustering parameters, are shown in Fig. \ref{fig:SD90vsRuntime}. We compare with the results for the optimal parameter choice for Adaptive Clustering.

The Density Clustering set-up chosen in \cite{Steltner:2020hfd} with a first-stage follow-up runtime-cost of $\leq$ 1 ATLAS-day is $\approx 10\%$ more sensitive than the Adaptive Clustering set-up at the same computing cost.  In continuous gravitational wave searches a $10\%$ improvement, solely due to a better search method, is a big gain. 

Perhaps more immediately impressive is the fact that at fixed sensitivity, Density Clustering enables follow-ups that are a factor of $\gtrsim 30$ faster than previous methods. 

This gain can be re-invested in deeper follow-ups by using a lower $\threshIter$, albeit the gain in practice is limited by the steep increase in computing cost for $\threshIter \lesssim 4$. With a threshold $\threshIter = -3.7$ Density Clustering is able to process two orders of magnitude more candidates than with a threshold $\threshIter = 4$, whereas Adaptive Clustering could not be used at all.

The performance of the Adaptive Clustering was characterised in \cite{Singh:2017kss} by the detection efficiency and the noise rejection $\NR$ defined as 
\begin{equation}
	\NR:=1-\frac{N_{out}}{N_{in}},
	\label{eq:noiseRejection}
\end{equation}
where $N_{in}$ is the number of candidates above the threshold $\threshIter$ and $N_{out}$ is the number of seeds produced by the clustering procedure. 	

With a threshold of $\threshIter \geq 4$ Adaptive Clustering and Density Clustering achieve similar performance with $\NR\geq 99\%$ and detection efficiencies above $98\%$. At lower thresholds, Adaptive Clustering does not converge in weeks of runtime, indicating that the method struggles to identify over densities due to faint signals. Density Clustering, instead, can probe threshold values as low as $-3.7$, still achieving $\NR \geq \NRDeClPercent\%$ and attaining a very respectable detection efficiency (now at the $85\%$ level) on a set that includes very faint signals with detection statistic values $\in [-3.7,4]$, which are much harder to find. 

\section{Conclusion and outlook}\label{sec:conclusion}

We have presented a new, fast and efficient clustering method - Density Clustering - for continuous gravitational wave search post-processing. 

Density Clustering works by identifying over-densities of candidates in parameter space: clusters are purely build on candidates' closeness to each other and the detection statistic value is nearly irrelevant. This result may be somewhat surprising because the detection statistic ranks  results based on the likeliness of they originating from a signal. However some of our faintest - but still recoverable - signals show detection statistic values at which there are thousands to millions of louder candidates purely from noise. Our results show that in this regime over-densities are a better detection criterion than the significance given by the detection statistic value alone, even in Gaussian noise. This is probably due to the fact that the search is a semi-coherent search.

The over-densities are uncovered by binning the parameter space and this is performed in one pass instead of the previously employed slower iterative procedures. The clustering step is thus largely independent of the number of input candidates, and this allows to process orders of magnitude more candidates with comparable computing resources, probing deeper into the noise. 

Until now Einstein@Home searches have returned about $\mathO{10^4}$ candidates per work-unit (e.g. \cite{Steltner:2020hfd}), which was more than adequate for what previous clustering algorithms could process. Density Clustering can cluster orders of magnitude more candidates, which means that more results can be inspected, allowing to recover fainter signals in upcoming searches.	

The previous clustering method, Adaptive Clustering, assumes compact over-densities, whereas signals typically present $X$- or $Y$-shaped over-densities which are hard to capture (and practically impossible to predict). Density Clustering is agnostic about the shape of the over-densities and for this reason it is significantly more effective at identifying even very weak signals.

A different approach of using machine learning for clustering was developed and applied to the Einstein@Home O2 all-sky dataset in  \cite{Beheshtipour:2020zhb, Beheshtipour:2020nko}. They cluster in $f, \fdot$ and achieve better sensitivity depths at fixed false alarms, but lack in sky localization to the point of clustering together candidates from ``seemingly unrelated sky positions'' \cite{Beheshtipour:2020nko}. This means that a follow-up would entail searching over the whole sky, whereas Density Clustering restricts the sky position to a patch of $\sim9\%$ to $0.01\%$ of the full sky, depending on the frequency, between $\sim \SI{20}{\hertz}$ to $\SI{600}{\hertz}$, respectively. Even with the smaller uncertainties in $f,\fdot$ and only half the false alarms \cite{Beheshtipour:2020nko}, the computational cost of their approach is higher by one order of magnitude compared to Density Clustering. They propose to generalize to include sky, and the results will be interesting to see.

Clustering is not a problem unique to gravitational wave astronomy, and a number of generic clustering methods exist. For example \emph{k}-means \cite{kmeans}, is a clustering method widely used in a variety of applications including signal-, image- and text-processing, health, cyber security, machine learning and big data \cite{electronics9081295}. It works based on minimising the cluster-occupants' distance to the cluster center. Limitations of \emph{k}-means are that the number of clusters must be known a-priori and clusters are assumed to be roughly spherical and similar size. Density-based clustering applications exist: for example DBSCAN \cite{dbscan, dbscan2} and its many generalizations, like e.g. OPTICS \cite{optics} or HDBSCAN \cite{hdbscan}, identify over-densities generated by a minimum number of points within a given volume. They are however not suitable for the large number of points in our results, and they are not as efficient as Density Clustering on our data.

A major advantage of our approach is the versatility of the method. Density Clustering can cluster in any combination of dimensions, so it is easily extendable to e.g. third / higher order spindowns $\dddot{f}, ...$ or to the 5 additional orbital parameters for searches for neutron stars in binary systems. In these searches signal-template offsets in orbital parameters can be to some extent compensated by offsets in frequency- and derivative(s). This translates into correlations between different templates and results in more candidates due to the same root cause \cite{Singh:2019han}, making clustering all the more important. All-sky binary searches are computationally extremely expensive and so are the follow-ups. A first test of Density Clustering on the results-data from \cite{Covas:2022rfg} showed promising results within a few hours of clustering in 6 dimensions $\left(f, \alpha,\delta,\tau_\mathrm{asc}, P_b, a\right)$, showcasing the flexibility and ease of use of the method presented here.

\acknowledgments
This work has utilised the ATLAS cluster computing at MPI for Gravitational Physics Hannover. We thank Carsten Aulbert and Henning Fehrmann for their support. We use results from the Einstein@Home search \cite{Steltner:2020hfd} on LIGO data obtained for that search from the Gravitational Wave Open Science Center (gw-openscience.org). We thank again the LIGO-Virgo-KAGRA Collaboration for this service, and LIGO for producing that data.
\newpage
\bibliography{paperBibDeCl} \bibliographystyle{plain}

\begin{thebibliography}{27}%
\makeatletter
\providecommand \@ifxundefined [1]{%
 \@ifx{#1\undefined}
}%
\providecommand \@ifnum [1]{%
 \ifnum #1\expandafter \@firstoftwo
 \else \expandafter \@secondoftwo
 \fi
}%
\providecommand \@ifx [1]{%
 \ifx #1\expandafter \@firstoftwo
 \else \expandafter \@secondoftwo
 \fi
}%
\providecommand \natexlab [1]{#1}%
\providecommand \enquote  [1]{``#1''}%
\providecommand \bibnamefont  [1]{#1}%
\providecommand \bibfnamefont [1]{#1}%
\providecommand \citenamefont [1]{#1}%
\providecommand \href@noop [0]{\@secondoftwo}%
\providecommand \href [0]{\begingroup \@sanitize@url \@href}%
\providecommand \@href[1]{\@@startlink{#1}\@@href}%
\providecommand \@@href[1]{\endgroup#1\@@endlink}%
\providecommand \@sanitize@url [0]{\catcode `\\12\catcode `\$12\catcode
  `\&12\catcode `\#12\catcode `\^12\catcode `\_12\catcode `\%12\relax}%
\providecommand \@@startlink[1]{}%
\providecommand \@@endlink[0]{}%
\providecommand \url  [0]{\begingroup\@sanitize@url \@url }%
\providecommand \@url [1]{\endgroup\@href {#1}{\urlprefix }}%
\providecommand \urlprefix  [0]{URL }%
\providecommand \Eprint [0]{\href }%
\providecommand \doibase [0]{https://doi.org/}%
\providecommand \selectlanguage [0]{\@gobble}%
\providecommand \bibinfo  [0]{\@secondoftwo}%
\providecommand \bibfield  [0]{\@secondoftwo}%
\providecommand \translation [1]{[#1]}%
\providecommand \BibitemOpen [0]{}%
\providecommand \bibitemStop [0]{}%
\providecommand \bibitemNoStop [0]{.\EOS\space}%
\providecommand \EOS [0]{\spacefactor3000\relax}%
\providecommand \BibitemShut  [1]{\csname bibitem#1\endcsname}%
\let\auto@bib@innerbib\@empty
\bibitem [{\citenamefont {Lasky}(2015)}]{Lasky:2015uia}%
  \BibitemOpen
  \bibfield  {author} {\bibinfo {author} {\bibfnamefont {P.~D.}\ \bibnamefont
  {Lasky}},\ }\bibfield  {title} {\bibinfo {title} {{Gravitational Waves from
  Neutron Stars: A Review}},\ }\href {https://doi.org/10.1017/pasa.2015.35}
  {\bibfield  {journal} {\bibinfo  {journal} {Publ. Astron. Soc. Austral.}\
  }\textbf {\bibinfo {volume} {32}},\ \bibinfo {pages} {e034} (\bibinfo {year}
  {2015})}\BibitemShut {NoStop}%
\bibitem [{\citenamefont {Owen}\ \emph {et~al.}(1998)\citenamefont {Owen},
  \citenamefont {Lindblom}, \citenamefont {Cutler}, \citenamefont {Schutz},
  \citenamefont {Vecchio},\ and\ \citenamefont {Andersson}}]{Owen:1998xg}%
  \BibitemOpen
  \bibfield  {author} {\bibinfo {author} {\bibfnamefont {B.~J.}\ \bibnamefont
  {Owen}}, \bibinfo {author} {\bibfnamefont {L.}~\bibnamefont {Lindblom}},
  \bibinfo {author} {\bibfnamefont {C.}~\bibnamefont {Cutler}}, \bibinfo
  {author} {\bibfnamefont {B.~F.}\ \bibnamefont {Schutz}}, \bibinfo {author}
  {\bibfnamefont {A.}~\bibnamefont {Vecchio}},\ and\ \bibinfo {author}
  {\bibfnamefont {N.}~\bibnamefont {Andersson}},\ }\bibfield  {title} {\bibinfo
  {title} {{Gravitational waves from hot young rapidly rotating neutron
  stars}},\ }\href {https://doi.org/10.1103/PhysRevD.58.084020} {\bibfield
  {journal} {\bibinfo  {journal} {Phys. Rev. D}\ }\textbf {\bibinfo {volume}
  {58}},\ \bibinfo {pages} {084020} (\bibinfo {year} {1998})}\BibitemShut
  {NoStop}%
\bibitem [{\citenamefont {Horowitz}\ and\ \citenamefont
  {Reddy}(2019)}]{Horowitz:2019aim}%
  \BibitemOpen
  \bibfield  {author} {\bibinfo {author} {\bibfnamefont {C.~J.}\ \bibnamefont
  {Horowitz}}\ and\ \bibinfo {author} {\bibfnamefont {S.}~\bibnamefont
  {Reddy}},\ }\bibfield  {title} {\bibinfo {title} {{Gravitational Waves from
  Compact Dark Objects in Neutron Stars}},\ }\href
  {https://doi.org/10.1103/PhysRevLett.122.071102} {\bibfield  {journal}
  {\bibinfo  {journal} {Phys. Rev. Lett.}\ }\textbf {\bibinfo {volume} {122}},\
  \bibinfo {pages} {071102} (\bibinfo {year} {2019})}\BibitemShut {NoStop}%
\bibitem [{\citenamefont {Horowitz}\ \emph {et~al.}(2020)\citenamefont
  {Horowitz}, \citenamefont {Papa},\ and\ \citenamefont
  {Reddy}}]{Horowitz:2019pru}%
  \BibitemOpen
  \bibfield  {author} {\bibinfo {author} {\bibfnamefont {C.}~\bibnamefont
  {Horowitz}}, \bibinfo {author} {\bibfnamefont {M.}~\bibnamefont {Papa}},\
  and\ \bibinfo {author} {\bibfnamefont {S.}~\bibnamefont {Reddy}},\ }\bibfield
   {title} {\bibinfo {title} {{Gravitational waves from compact dark matter
  objects in the solar system}},\ }\href
  {https://doi.org/10.1016/j.physletb.2019.135072} {\bibfield  {journal}
  {\bibinfo  {journal} {Phys. Lett. B}\ }\textbf {\bibinfo {volume} {800}},\
  \bibinfo {pages} {135072} (\bibinfo {year} {2020})}\BibitemShut {NoStop}%
\bibitem [{\citenamefont {Arvanitaki}\ \emph {et~al.}(2015)\citenamefont
  {Arvanitaki}, \citenamefont {Baryakhtar},\ and\ \citenamefont
  {Huang}}]{Arvanitaki:2014wva}%
  \BibitemOpen
  \bibfield  {author} {\bibinfo {author} {\bibfnamefont {A.}~\bibnamefont
  {Arvanitaki}}, \bibinfo {author} {\bibfnamefont {M.}~\bibnamefont
  {Baryakhtar}},\ and\ \bibinfo {author} {\bibfnamefont {X.}~\bibnamefont
  {Huang}},\ }\bibfield  {title} {\bibinfo {title} {{Discovering the QCD Axion
  with Black Holes and Gravitational Waves}},\ }\href
  {https://doi.org/10.1103/PhysRevD.91.084011} {\bibfield  {journal} {\bibinfo
  {journal} {Phys. Rev. D}\ }\textbf {\bibinfo {volume} {91}},\ \bibinfo
  {pages} {084011} (\bibinfo {year} {2015})}\BibitemShut {NoStop}%
\bibitem [{\citenamefont {Zhu}\ \emph {et~al.}(2020)\citenamefont {Zhu},
  \citenamefont {Baryakhtar}, \citenamefont {Papa}, \citenamefont {Tsuna},
  \citenamefont {Kawanaka},\ and\ \citenamefont {Eggenstein}}]{Zhu:2020tht}%
  \BibitemOpen
  \bibfield  {author} {\bibinfo {author} {\bibfnamefont {S.~J.}\ \bibnamefont
  {Zhu}}, \bibinfo {author} {\bibfnamefont {M.}~\bibnamefont {Baryakhtar}},
  \bibinfo {author} {\bibfnamefont {M.~A.}\ \bibnamefont {Papa}}, \bibinfo
  {author} {\bibfnamefont {D.}~\bibnamefont {Tsuna}}, \bibinfo {author}
  {\bibfnamefont {N.}~\bibnamefont {Kawanaka}},\ and\ \bibinfo {author}
  {\bibfnamefont {H.-B.}\ \bibnamefont {Eggenstein}},\ }\bibfield  {title}
  {\bibinfo {title} {{Characterizing the continuous gravitational-wave signal
  from boson clouds around Galactic isolated black holes}},\ }\href
  {https://doi.org/10.1103/PhysRevD.102.063020} {\bibfield  {journal} {\bibinfo
   {journal} {Phys. Rev. D}\ }\textbf {\bibinfo {volume} {102}},\ \bibinfo
  {pages} {063020} (\bibinfo {year} {2020})}\BibitemShut {NoStop}%
\bibitem [{\citenamefont {Papa}\ \emph {et~al.}(2016)\citenamefont {Papa} \emph
  {et~al.}}]{Papa:2016cwb}%
  \BibitemOpen
  \bibfield  {author} {\bibinfo {author} {\bibfnamefont {M.~A.}\ \bibnamefont
  {Papa}} \emph {et~al.},\ }\bibfield  {title} {\bibinfo {title} {{Hierarchical
  follow-up of subthreshold candidates of an all-sky Einstein@Home search for
  continuous gravitational waves on LIGO sixth science run data}},\ }\href
  {https://doi.org/10.1103/PhysRevD.94.122006} {\bibfield  {journal} {\bibinfo
  {journal} {Phys. Rev. D}\ }\textbf {\bibinfo {volume} {94}},\ \bibinfo
  {pages} {122006} (\bibinfo {year} {2016})}\BibitemShut {NoStop}%
\bibitem [{\citenamefont {Abbott}\ \emph {et~al.}(2017)\citenamefont {Abbott}
  \emph {et~al.}}]{Abbott:2017pqa}%
  \BibitemOpen
  \bibfield  {author} {\bibinfo {author} {\bibfnamefont {B.~P.}\ \bibnamefont
  {Abbott}} \emph {et~al.} (\bibinfo {collaboration} {LIGO Scientific,
  Virgo}),\ }\bibfield  {title} {\bibinfo {title} {{First low-frequency
  Einstein@Home all-sky search for continuous gravitational waves in Advanced
  LIGO data}},\ }\href {https://doi.org/10.1103/PhysRevD.96.122004} {\bibfield
  {journal} {\bibinfo  {journal} {Phys. Rev.}\ }\textbf {\bibinfo {volume}
  {D96}},\ \bibinfo {pages} {122004} (\bibinfo {year} {2017})}\BibitemShut
  {NoStop}%
\bibitem [{\citenamefont {Steltner}\ \emph {et~al.}(2021)\citenamefont
  {Steltner}, \citenamefont {Papa}, \citenamefont {Eggenstein}, \citenamefont
  {Allen}, \citenamefont {Dergachev}, \citenamefont {Prix}, \citenamefont
  {Machenschalk}, \citenamefont {Walsh}, \citenamefont {Zhu},\ and\
  \citenamefont {Kwang}}]{Steltner:2020hfd}%
  \BibitemOpen
  \bibfield  {author} {\bibinfo {author} {\bibfnamefont {B.}~\bibnamefont
  {Steltner}}, \bibinfo {author} {\bibfnamefont {M.~A.}\ \bibnamefont {Papa}},
  \bibinfo {author} {\bibfnamefont {H.~B.}\ \bibnamefont {Eggenstein}},
  \bibinfo {author} {\bibfnamefont {B.}~\bibnamefont {Allen}}, \bibinfo
  {author} {\bibfnamefont {V.}~\bibnamefont {Dergachev}}, \bibinfo {author}
  {\bibfnamefont {R.}~\bibnamefont {Prix}}, \bibinfo {author} {\bibfnamefont
  {B.}~\bibnamefont {Machenschalk}}, \bibinfo {author} {\bibfnamefont
  {S.}~\bibnamefont {Walsh}}, \bibinfo {author} {\bibfnamefont {S.~J.}\
  \bibnamefont {Zhu}},\ and\ \bibinfo {author} {\bibfnamefont {S.}~\bibnamefont
  {Kwang}},\ }\bibfield  {title} {\bibinfo {title} {{Einstein@Home All-sky
  Search for Continuous Gravitational Waves in LIGO O2 Public Data}},\ }\href
  {https://doi.org/10.3847/1538-4357/abc7c9} {\bibfield  {journal} {\bibinfo
  {journal} {Astrophys. J.}\ }\textbf {\bibinfo {volume} {909}},\ \bibinfo
  {pages} {79} (\bibinfo {year} {2021})}\BibitemShut {NoStop}%
\bibitem [{\citenamefont {BOINC}(2020)}]{Boinc1}%
  \BibitemOpen
  \bibfield  {author} {\bibinfo {author} {\bibnamefont {BOINC}},\ }\href@noop
  {} {}\bibinfo {howpublished} {\url{http://boinc.berkeley.edu/}} (\bibinfo
  {year} {2020})\BibitemShut {NoStop}%
\bibitem [{\citenamefont {{Anderson}}(2004)}]{Boinc2}%
  \BibitemOpen
  \bibfield  {author} {\bibinfo {author} {\bibfnamefont {D.~P.}\ \bibnamefont
  {{Anderson}}},\ }\bibfield  {title} {\bibinfo {title} {{BOINC: A System for
  Public-Resource Computing and Storage}},\ }in\ \href@noop {} {\emph {\bibinfo
  {booktitle} {Proceedings of the Fifth IEEE/ACM International Workshop on Grid
  Computing (GRID04)}}}\ (\bibinfo {year} {2004})\ pp.\ \bibinfo {pages}
  {4--10}\BibitemShut {NoStop}%
\bibitem [{\citenamefont {{Anderson}}\ \emph {et~al.}(2006)\citenamefont
  {{Anderson}}, \citenamefont {{Christensen}},\ and\ \citenamefont
  {{Allen}}}]{Boinc3}%
  \BibitemOpen
  \bibfield  {author} {\bibinfo {author} {\bibfnamefont {D.~P.}\ \bibnamefont
  {{Anderson}}}, \bibinfo {author} {\bibfnamefont {C.}~\bibnamefont
  {{Christensen}}},\ and\ \bibinfo {author} {\bibfnamefont {B.}~\bibnamefont
  {{Allen}}},\ }\bibfield  {title} {\bibinfo {title} {{Designing a Runtime
  System for Volunteer Computing}},\ }in\ \href@noop {} {\emph {\bibinfo
  {booktitle} {Proceedings of the 2006 ACM/IEEE conference on
  Supercomputing}}}\ (\bibinfo {year} {2006})\ pp.\ \bibinfo {pages}
  {126--136}\BibitemShut {NoStop}%
\bibitem [{\citenamefont {Aasi}\ \emph {et~al.}(2013)\citenamefont {Aasi} \emph
  {et~al.}}]{Aasi:2013jya}%
  \BibitemOpen
  \bibfield  {author} {\bibinfo {author} {\bibfnamefont {J.}~\bibnamefont
  {Aasi}} \emph {et~al.} (\bibinfo {collaboration} {LIGO Scientific, VIRGO}),\
  }\bibfield  {title} {\bibinfo {title} {{Directed search for continuous
  gravitational waves from the Galactic center}},\ }\href
  {https://doi.org/10.1103/PhysRevD.88.102002} {\bibfield  {journal} {\bibinfo
  {journal} {Phys. Rev. D}\ }\textbf {\bibinfo {volume} {88}},\ \bibinfo
  {pages} {102002} (\bibinfo {year} {2013})}\BibitemShut {NoStop}%
\bibitem [{\citenamefont {Behnke}\ \emph {et~al.}(2015)\citenamefont {Behnke},
  \citenamefont {Papa},\ and\ \citenamefont {Prix}}]{Behnke:2014tma}%
  \BibitemOpen
  \bibfield  {author} {\bibinfo {author} {\bibfnamefont {B.}~\bibnamefont
  {Behnke}}, \bibinfo {author} {\bibfnamefont {M.~A.}\ \bibnamefont {Papa}},\
  and\ \bibinfo {author} {\bibfnamefont {R.}~\bibnamefont {Prix}},\ }\bibfield
  {title} {\bibinfo {title} {{Postprocessing methods used in the search for
  continuous gravitational-wave signals from the Galactic Center}},\ }\href
  {https://doi.org/10.1103/PhysRevD.91.064007} {\bibfield  {journal} {\bibinfo
  {journal} {Phys. Rev. D}\ }\textbf {\bibinfo {volume} {91}},\ \bibinfo
  {pages} {064007} (\bibinfo {year} {2015})}\BibitemShut {NoStop}%
\bibitem [{\citenamefont {Singh}\ \emph {et~al.}(2017)\citenamefont {Singh},
  \citenamefont {Papa}, \citenamefont {Eggenstein},\ and\ \citenamefont
  {Walsh}}]{Singh:2017kss}%
  \BibitemOpen
  \bibfield  {author} {\bibinfo {author} {\bibfnamefont {A.}~\bibnamefont
  {Singh}}, \bibinfo {author} {\bibfnamefont {M.~A.}\ \bibnamefont {Papa}},
  \bibinfo {author} {\bibfnamefont {H.-B.}\ \bibnamefont {Eggenstein}},\ and\
  \bibinfo {author} {\bibfnamefont {S.}~\bibnamefont {Walsh}},\ }\bibfield
  {title} {\bibinfo {title} {{Adaptive clustering procedure for continuous
  gravitational wave searches}},\ }\href
  {https://doi.org/10.1103/PhysRevD.96.082003} {\bibfield  {journal} {\bibinfo
  {journal} {Phys. Rev. D}\ }\textbf {\bibinfo {volume} {96}},\ \bibinfo
  {pages} {082003} (\bibinfo {year} {2017})}\BibitemShut {NoStop}%
\bibitem [{\citenamefont {Beheshtipour}\ and\ \citenamefont
  {Papa}(2020)}]{Beheshtipour:2020zhb}%
  \BibitemOpen
  \bibfield  {author} {\bibinfo {author} {\bibfnamefont {B.}~\bibnamefont
  {Beheshtipour}}\ and\ \bibinfo {author} {\bibfnamefont {M.~A.}\ \bibnamefont
  {Papa}},\ }\bibfield  {title} {\bibinfo {title} {{Deep learning for
  clustering of continuous gravitational wave candidates}},\ }\href
  {https://doi.org/10.1103/PhysRevD.101.064009} {\bibfield  {journal} {\bibinfo
   {journal} {Phys. Rev. D}\ }\textbf {\bibinfo {volume} {101}},\ \bibinfo
  {pages} {064009} (\bibinfo {year} {2020})}\BibitemShut {NoStop}%
\bibitem [{\citenamefont {Beheshtipour}\ and\ \citenamefont
  {Papa}(2021)}]{Beheshtipour:2020nko}%
  \BibitemOpen
  \bibfield  {author} {\bibinfo {author} {\bibfnamefont {B.}~\bibnamefont
  {Beheshtipour}}\ and\ \bibinfo {author} {\bibfnamefont {M.~A.}\ \bibnamefont
  {Papa}},\ }\bibfield  {title} {\bibinfo {title} {{Deep learning for
  clustering of continuous gravitational wave candidates II: identification of
  low-SNR candidates}},\ }\href {https://doi.org/10.1103/PhysRevD.103.064027}
  {\bibfield  {journal} {\bibinfo  {journal} {Phys. Rev. D}\ }\textbf {\bibinfo
  {volume} {103}},\ \bibinfo {pages} {064027} (\bibinfo {year}
  {2021})}\BibitemShut {NoStop}%
\bibitem [{\citenamefont {Vallisneri}\ \emph {et~al.}(2015)\citenamefont
  {Vallisneri}, \citenamefont {Kanner}, \citenamefont {Williams}, \citenamefont
  {Weinstein},\ and\ \citenamefont {Stephens}}]{Vallisneri:2014vxa}%
  \BibitemOpen
  \bibfield  {author} {\bibinfo {author} {\bibfnamefont {M.}~\bibnamefont
  {Vallisneri}}, \bibinfo {author} {\bibfnamefont {J.}~\bibnamefont {Kanner}},
  \bibinfo {author} {\bibfnamefont {R.}~\bibnamefont {Williams}}, \bibinfo
  {author} {\bibfnamefont {A.}~\bibnamefont {Weinstein}},\ and\ \bibinfo
  {author} {\bibfnamefont {B.}~\bibnamefont {Stephens}},\ }\bibfield  {title}
  {\bibinfo {title} {{The LIGO Open Science Center}},\ }\bibfield  {booktitle}
  {\emph {\bibinfo {booktitle} {{Proceedings, 10th International LISA
  Symposium: Gainesville, Florida, USA, May 18-23, 2014}}},\ }\href
  {https://doi.org/10.1088/1742-6596/610/1/012021} {\bibfield  {journal}
  {\bibinfo  {journal} {J. Phys. Conf. Ser.}\ }\textbf {\bibinfo {volume}
  {610}},\ \bibinfo {pages} {012021} (\bibinfo {year} {2015})}\BibitemShut
  {NoStop}%
\bibitem [{\citenamefont {Keitel}(2016)}]{Keitel:2015ova}%
  \BibitemOpen
  \bibfield  {author} {\bibinfo {author} {\bibfnamefont {D.}~\bibnamefont
  {Keitel}},\ }\bibfield  {title} {\bibinfo {title} {{Robust semicoherent
  searches for continuous gravitational waves with noise and signal models
  including hours to days long transients}},\ }\href
  {https://doi.org/10.1103/PhysRevD.93.084024} {\bibfield  {journal} {\bibinfo
  {journal} {Phys. Rev.}\ }\textbf {\bibinfo {volume} {D93}},\ \bibinfo {pages}
  {084024} (\bibinfo {year} {2016})},\ \Eprint
  {https://arxiv.org/abs/1509.02398} {1509.02398} \BibitemShut {NoStop}%
\bibitem [{\citenamefont {Lloyd}(1982)}]{kmeans}%
  \BibitemOpen
  \bibfield  {author} {\bibinfo {author} {\bibfnamefont {S.}~\bibnamefont
  {Lloyd}},\ }\bibfield  {title} {\bibinfo {title} {Least squares quantization
  in pcm},\ }\href {https://doi.org/10.1109/TIT.1982.1056489} {\bibfield
  {journal} {\bibinfo  {journal} {IEEE Transactions on Information Theory}\
  }\textbf {\bibinfo {volume} {28}},\ \bibinfo {pages} {129} (\bibinfo {year}
  {1982})}\BibitemShut {NoStop}%
\bibitem [{\citenamefont {Ahmed}\ \emph {et~al.}(2020)\citenamefont {Ahmed},
  \citenamefont {Seraj},\ and\ \citenamefont {Islam}}]{electronics9081295}%
  \BibitemOpen
  \bibfield  {author} {\bibinfo {author} {\bibfnamefont {M.}~\bibnamefont
  {Ahmed}}, \bibinfo {author} {\bibfnamefont {R.}~\bibnamefont {Seraj}},\ and\
  \bibinfo {author} {\bibfnamefont {S.~M.~S.}\ \bibnamefont {Islam}},\
  }\bibfield  {title} {\bibinfo {title} {The k-means algorithm: A comprehensive
  survey and performance evaluation},\ }\bibfield  {journal} {\bibinfo
  {journal} {Electronics}\ }\textbf {\bibinfo {volume} {9}},\ \href
  {https://doi.org/10.3390/electronics9081295} {10.3390/electronics9081295}
  (\bibinfo {year} {2020})\BibitemShut {NoStop}%
\bibitem [{\citenamefont {Ester}\ \emph {et~al.}(1996)\citenamefont {Ester},
  \citenamefont {Kriegel}, \citenamefont {Sander},\ and\ \citenamefont
  {Xu}}]{dbscan}%
  \BibitemOpen
  \bibfield  {author} {\bibinfo {author} {\bibfnamefont {M.}~\bibnamefont
  {Ester}}, \bibinfo {author} {\bibfnamefont {H.-P.}\ \bibnamefont {Kriegel}},
  \bibinfo {author} {\bibfnamefont {J.}~\bibnamefont {Sander}},\ and\ \bibinfo
  {author} {\bibfnamefont {X.}~\bibnamefont {Xu}},\ }\bibfield  {title}
  {\bibinfo {title} {A density-based algorithm for discovering clusters in
  large spatial databases with noise},\ }in\ \href@noop {} {\emph {\bibinfo
  {booktitle} {Proceedings of the Second International Conference on Knowledge
  Discovery and Data Mining}}},\ \bibinfo {series and number} {KDD'96}\
  (\bibinfo  {publisher} {AAAI Press},\ \bibinfo {year} {1996})\ p.\ \bibinfo
  {pages} {226–231}\BibitemShut {NoStop}%
\bibitem [{\citenamefont {Schubert}\ \emph {et~al.}(2017)\citenamefont
  {Schubert}, \citenamefont {Sander}, \citenamefont {Ester}, \citenamefont
  {Kriegel},\ and\ \citenamefont {Xu}}]{dbscan2}%
  \BibitemOpen
  \bibfield  {author} {\bibinfo {author} {\bibfnamefont {E.}~\bibnamefont
  {Schubert}}, \bibinfo {author} {\bibfnamefont {J.}~\bibnamefont {Sander}},
  \bibinfo {author} {\bibfnamefont {M.}~\bibnamefont {Ester}}, \bibinfo
  {author} {\bibfnamefont {H.~P.}\ \bibnamefont {Kriegel}},\ and\ \bibinfo
  {author} {\bibfnamefont {X.}~\bibnamefont {Xu}},\ }\bibfield  {title}
  {\bibinfo {title} {Dbscan revisited, revisited: Why and how you should
  (still) use dbscan},\ }\bibfield  {journal} {\bibinfo  {journal} {ACM Trans.
  Database Syst.}\ }\textbf {\bibinfo {volume} {42}},\ \href
  {https://doi.org/10.1145/3068335} {10.1145/3068335} (\bibinfo {year}
  {2017})\BibitemShut {NoStop}%
\bibitem [{\citenamefont {Ankerst}\ \emph {et~al.}(1999)\citenamefont
  {Ankerst}, \citenamefont {Breunig}, \citenamefont {Kriegel},\ and\
  \citenamefont {Sander}}]{optics}%
  \BibitemOpen
  \bibfield  {author} {\bibinfo {author} {\bibfnamefont {M.}~\bibnamefont
  {Ankerst}}, \bibinfo {author} {\bibfnamefont {M.~M.}\ \bibnamefont
  {Breunig}}, \bibinfo {author} {\bibfnamefont {H.-P.}\ \bibnamefont
  {Kriegel}},\ and\ \bibinfo {author} {\bibfnamefont {J.}~\bibnamefont
  {Sander}},\ }\bibfield  {title} {\bibinfo {title} {Optics: Ordering points to
  identify the clustering structure},\ }\href
  {https://doi.org/10.1145/304181.304187} {\bibfield  {journal} {\bibinfo
  {journal} {SIGMOD Rec.}\ }\textbf {\bibinfo {volume} {28}},\ \bibinfo {pages}
  {49–60} (\bibinfo {year} {1999})}\BibitemShut {NoStop}%
\bibitem [{\citenamefont {Campello}\ \emph {et~al.}(2013)\citenamefont
  {Campello}, \citenamefont {Moulavi},\ and\ \citenamefont {Sander}}]{hdbscan}%
  \BibitemOpen
  \bibfield  {author} {\bibinfo {author} {\bibfnamefont {R.~J. G.~B.}\
  \bibnamefont {Campello}}, \bibinfo {author} {\bibfnamefont {D.}~\bibnamefont
  {Moulavi}},\ and\ \bibinfo {author} {\bibfnamefont {J.}~\bibnamefont
  {Sander}},\ }\bibfield  {title} {\bibinfo {title} {Density-based clustering
  based on hierarchical density estimates},\ }in\ \href@noop {} {\emph
  {\bibinfo {booktitle} {Advances in Knowledge Discovery and Data Mining}}},\
  \bibinfo {editor} {edited by\ \bibinfo {editor} {\bibfnamefont
  {J.}~\bibnamefont {Pei}}, \bibinfo {editor} {\bibfnamefont {V.~S.}\
  \bibnamefont {Tseng}}, \bibinfo {editor} {\bibfnamefont {L.}~\bibnamefont
  {Cao}}, \bibinfo {editor} {\bibfnamefont {H.}~\bibnamefont {Motoda}},\ and\
  \bibinfo {editor} {\bibfnamefont {G.}~\bibnamefont {Xu}}}\ (\bibinfo
  {publisher} {Springer Berlin Heidelberg},\ \bibinfo {address} {Berlin,
  Heidelberg},\ \bibinfo {year} {2013})\ pp.\ \bibinfo {pages}
  {160--172}\BibitemShut {NoStop}%
\bibitem [{\citenamefont {Singh}\ \emph {et~al.}(2019)\citenamefont {Singh},
  \citenamefont {Papa},\ and\ \citenamefont {Dergachev}}]{Singh:2019han}%
  \BibitemOpen
  \bibfield  {author} {\bibinfo {author} {\bibfnamefont {A.}~\bibnamefont
  {Singh}}, \bibinfo {author} {\bibfnamefont {M.~A.}\ \bibnamefont {Papa}},\
  and\ \bibinfo {author} {\bibfnamefont {V.}~\bibnamefont {Dergachev}},\
  }\bibfield  {title} {\bibinfo {title} {{Characterizing the sensitivity of
  isolated continuous gravitational wave searches to binary orbits}},\ }\href
  {https://doi.org/10.1103/PhysRevD.100.024058} {\bibfield  {journal} {\bibinfo
   {journal} {Phys. Rev. D}\ }\textbf {\bibinfo {volume} {100}},\ \bibinfo
  {pages} {024058} (\bibinfo {year} {2019})}\BibitemShut {NoStop}%
\bibitem [{\citenamefont {Covas}\ \emph {et~al.}(2022)\citenamefont {Covas},
  \citenamefont {Papa}, \citenamefont {Prix},\ and\ \citenamefont
  {Owen}}]{Covas:2022rfg}%
  \BibitemOpen
  \bibfield  {author} {\bibinfo {author} {\bibfnamefont {P.~B.}\ \bibnamefont
  {Covas}}, \bibinfo {author} {\bibfnamefont {M.~A.}\ \bibnamefont {Papa}},
  \bibinfo {author} {\bibfnamefont {R.}~\bibnamefont {Prix}},\ and\ \bibinfo
  {author} {\bibfnamefont {B.~J.}\ \bibnamefont {Owen}},\ }\bibfield  {title}
  {\bibinfo {title} {{Constraints on r-modes and Mountains on Millisecond
  Neutron Stars in Binary Systems}},\ }\href
  {https://doi.org/10.3847/2041-8213/ac62d7} {\bibfield  {journal} {\bibinfo
  {journal} {Astrophys. J. Lett.}\ }\textbf {\bibinfo {volume} {929}},\
  \bibinfo {pages} {L19} (\bibinfo {year} {2022})}\BibitemShut {NoStop}%
\end{thebibliography}%


\end{document}